\documentclass[
reprint,
amsmath,amssymb,
aps,
]{revtex4-2}

\usepackage{tabularray}
\UseTblrLibrary{booktabs}
\UseTblrLibrary{booktabs}
\usepackage{import}
\usepackage{array}
\usepackage{adjustbox}
\usepackage{graphicx}
\usepackage{dcolumn}
\usepackage{bm}
\usepackage{hyperref}
\usepackage{xcolor}

\usepackage{verbatim}    
\usepackage{tikz} 
\usepackage{mathtools}
\usepackage{comment}

\makeatletter
\newcommand{\cbigoplus}{\DOTSB\cbigoplus@\slimits@}

\DeclarePairedDelimiterX\braket[2]{\langle}{\rangle}{#1\,\delimsize\vert\,\mathopen{}#2}

\DeclareMathOperator{\Tr}{Tr}

\begin{document}
	
	\preprint{APS/123-QED}
	
	\title{Boosting Gaussian Boson Sampling using Optical Parametric Amplification Networks }
	
	
	\author{Yukuan Zhao}
	\affiliation{Laboratory of Quantum Information, University of Science and Technology of China, Hefei 230026, China}
	\affiliation{School of Physical Sciences, University of Science and Technology of China, Hefei 230026, China}
	\author{Xiao-Ye Xu}
	\email{xuxiaoye@ustc.edu.cn}
	\affiliation{Laboratory of Quantum Information, University of Science and Technology of China, Hefei 230026, China}
	\affiliation{Hefei National Laboratory, University of Science and Technology of China, Hefei 230088, China}
    \affiliation{Anhui Province Key Laboratory of Quantum Network, University of Science and Technology of China, Hefei 230026, China}
	\author{Chuan-Feng Li}
	\email{cfli@ustc.edu.cn}
	\author{Guang-Can Guo}
	\affiliation{Laboratory of Quantum Information, University of Science and Technology of China, Hefei 230026, China}
	\affiliation{Hefei National Laboratory, University of Science and Technology of China, Hefei 230088, China}
        \affiliation{Anhui Province Key Laboratory of Quantum Network, University of Science and Technology of China, Hefei 230026, China}

	
	\date{\today}
	
	\begin{abstract}
        Gaussian Boson Sampling (GBS) provides a route toward demonstrating quantum computational advantage. However, optical loss, which reduces the entanglement in the 	system, can render GBS results classically simulable. We propose a nonlinear photonic architecture based on optical parametric amplifiers (OPAs) arranged in an interferometer network. This active configuration amplifies quantum correlations within the circuit while preserving the \#P-hard Hafnian structure of the output probabilities. Using logarithmic negativity, we numerically show that entanglement scales linearly with both the OPA gain and network depth in the lossless limit, and maintains linear scaling with the number of modes under realistic loss rate. These scaling behaviors suggest that classical simulation in lossy scenarios remains computationally intractable. The decomposition of the output into a core state and a random Gaussian displacement shows that SU(1,1) network outperforms SU(2) network when loss is present. Our results demonstrate that OPA-boosted GBS preserves computational hardness in noisy environments, offering a more effective implementations of near-term photonic quantum computers.
	\end{abstract}
	
	\maketitle
	
	\emph{Introduction.---}
    Quantum computational advantage occurs when a quantum device performs a task that cannot be efficiently simulated by any known classical algorithm. One of the most prominent proposals to demonstrate this phenomenon is \emph{Boson Sampling} (BS)~\cite{aaronson2010computational,Gard_2015}, in which $n$ indistinguishable photons interfere within a linear optical network composed of beam splitters and phase shifters. The output probabilities are determined by matrix permanents, a $\#P$-hard quantity, making exact or approximate classical simulation infeasible under widely accepted complexity-theoretic assumptions. In the ideal lossless case, only $n \sim 50$ single-photon events are expected to exceed the capabilities of the best classical algorithms~\cite{Neville2017}.  
    
    Early experiments relied on probabilistic single-photon sources based on spontaneous parametric down-conversion (SPDC)~\cite{Tillmann2013,doi:10.1126/science.1231692,doi:10.1126/science.1231440}, but were limited to a few photons—insufficient for demonstrating quantum advantage. Subsequent implementations using deterministic quantum-dot emitters have reached up to 20 photons~\cite{PhysRevLett.123.250503}. The development of \emph{Gaussian Boson Sampling} (GBS)~\cite{Hamilton_2017} overcame many of these limitations by replacing single photons with squeezed vacuum states, whose output probabilities are governed by Hafnians of covariance matrices, also $\#P$-hard to compute. This approach has enabled large-scale photonic demonstrations, notably Jiuzhang\cite{Zhong2020Jiuzhang1,Zhong2021Jiuzhang2,PhysRevLett.131.150601,liu2025robustquantumcomputationaladvantage} and the programmable Borealis processor~\cite{Madsen2022}, which detected hundreds of photons and exhibited quantum advantage. 
    
    Despite these advances, experimental imperfections—particularly photon loss and partial indistinguishability—severely impact the computational hardness of GBS. Photon loss reduces output-state entanglement and allows efficient classical simulations~\cite{PhysRevA.93.012335,PhysRevX.6.021039,Garc_a_Patr_n_2019,Oszmaniec_2018}. Tensor-network methods, including matrix product operator (MPO) and matrix product state (MPS) algorithms, have been shown to efficiently approximate GBS outputs whenever the entanglement entropy scales sublinearly, induced by photon loss, with system size~\cite{PhysRevA.104.022407,Oh2024}. In contrast, when entanglement scales linearly or faster with the number of modes, classical simulation requires exponential resources~\cite{PhysRevA.108.052604}. 


	\begin{figure*}
		\includegraphics[width=0.9\textwidth]{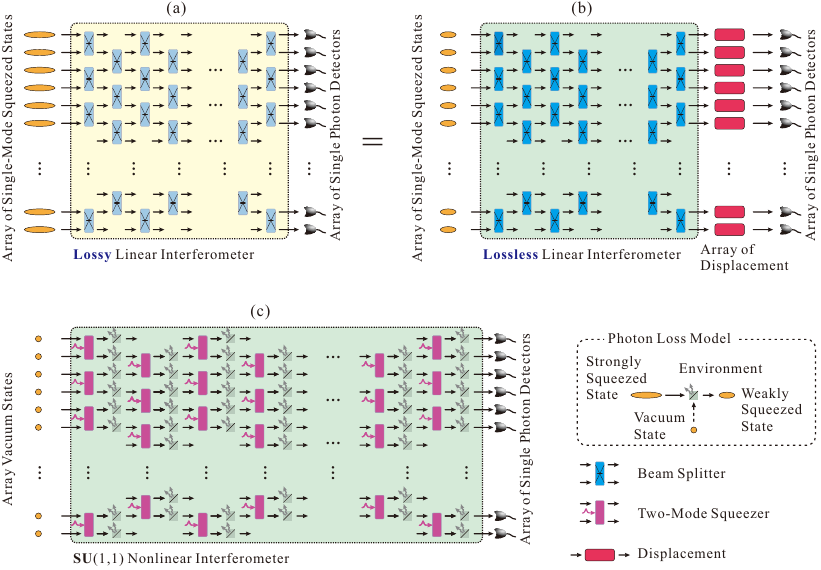}
		\caption{(a) depicts the model of a lossy GBS. The inputs are single-mode squeezed states (SMSS) going through a lossy linear interferometer composed of arranged beam splitters, and (a) is equivalent to (b), using weaker squeezed SMSS passing through a lossless linear interferometer followed by an array of displacement operators. (c) shows the proposed nonlinear interferometer where the beam splitters of GBS are replaced by OPAs described by elements of $\mathrm{SU(1,1)}$. Beam splitter between OPA layers are used to model loss. Instead of inputting the squeezed states, here an array of vacuum states is adopted as the source.}
		\label{fig:Model}
	\end{figure*}
	
	A key limitation of current GBS is that the entanglement originates solely from the nonclassicality of the input states and the array of interferometers acts as a passive entanglement shuffler\,\cite{Kim2002}. As illustrated in Fig.\,\ref{fig:Model}(a), standard GBS involves an array of single-mode squeezed states (SMSS) propagating through a linear optical network composed of beam splitters. 
	As the states traverse the network, photon loss inevitably occurs.
	This lossy configuration is statistically equivalent to one in which an array of less-squeezed SMSS inputs propagates through a lossless interferometer as shown in Fig\,\ref{fig:Model}(b)\,\cite{Oh2024}. 
	Photon loss reduces the entanglement within a sampling system, and, at present, there is no mechanism in linear interferometers to recover the entanglement lost to the environment. Moreover, current experimental platforms face substantial challenges in further minimizing optical loss, posing a bottleneck to scalability and performance improvements.

 In this work, we propose a boosted GBS scheme in which the interferometer array itself contributes additional nonclassicality to counteract entanglement degradation due to loss. While nonlinear extensions to BS have previously been considered to enhance expressivity \,\cite{Spagnolo2023}, the issue of loss remains unresolved. Our approach leverages the nonlinear interaction provided by optical parametric amplifiers (OPA)\,\cite{PhysRev.160.1076}, described by the $\mathrm{SU}(1,1)$ group\,\cite{PhysRevA.33.4033}, to mitigate the reduction of entanglement from photon loss. This transforms the optical network from a purely passive to an active system, enhancing entanglement without altering the underlying computational complexity class of the sampling problem, as we demonstrate in subsequent sections.  
In Appendix D, we present a quick view on the improvement of our proposed OPA-boosted scheme in comparison with the original Boson Sampling as well as its variants.
 Although, we focus on optical implementation, other physical implementations, such as, ion trap could also be viable\,\cite{PhysRevLett.112.050504, rblumel_1995}.

\emph{Theoretical model of the OPA network.---}
	As depicted in Fig.\ref{fig:Model}(c), the proposed GBS consists of a multilayer network of two-mode squeezers, implemented using second-order nonlinear OPAs. 
    These OPAs are arranged in a staggered configuration such that outputs from each layer serve as inputs to the next. With a sufficient number of layers, this architecture achieves full connectivity among all modes.
	To obtain enough gain or degree of squeezing, we consider using an ultrafast laser with pulse width of a few hundred femtoseconds. Fortunately, recent advances in frequency decoupling technology\,\cite{Harder2016} allow spectral uniformity between the two modes of each OPA. The laser pulses used to pump all the OPAs should be phase-locked to maintain the coherence of the entire system. This synchronized pumping technique has been adopted in continuous-variable quantum information to generate multimode squeezed states \cite{Patera2010, Averchenko2011}. \textcolor{blue}{The technology required for realizing this device is within reach.}
	
	To model photon loss, we introduce beam splitter layers between adjacent OPA layers, as shown in Fig.\,\ref{fig:Model}(c) and the inset. The reflecting channel of each beam splitter is discarded, while the photons that are successfully transmitted are directed to the next OPA layer. In an ideal network, the transmittance $t$ of the beam splitters is equal to one.
 
	We first present the final state for a lossless network, which is a multimode nonclassical Gaussian state. The nonlinear interaction introduced by the OPA is governed by the Hamiltonian $H=\hbar(\xi \hat{a}^{\dagger}\hat{b}^{\dagger}-\xi^{*}\hat{a}\hat{b})$, which results in a squeezing operation on the two modes defined by the creation  $a^\dagger$,$b^\dagger$ and annihilation operators $a$, $b$. 
	The complex number $\xi=-re^{i\theta}$ represents the squeezing parameter where $r$ determines the degree of squeezing and $\theta$ the squeezing angle.
	Furthermore, an $n$-mode Gaussian state can be characterized by a $2n \times 2n$ matrix \,\cite{RevModPhys.84.621} $\sigma$ with entries
	\begin{equation}
		\sigma_{ij}=\frac{\left\langle q_iq_j +q_jq_i\right\rangle}{2}-\left\langle q_i \right\rangle\left\langle q_j \right\rangle \label{covariance}
	\end{equation}
	where $q_i$ are elements of canonically conjugate quadrature components for $q=(x_1,x_2,...,x_n;p_1,p_2...,p_n)$. 

	This covariance matrix provides a complete description of the multimode Gaussian state and serves as the foundation for analyzing the system's behavior under the influence of OPAs as well as beam splitters.
	Let $\sigma$ be the covariance matrix associated with the input multimode Gaussian state. As the state evolves through the lossless OPA array, with transmittance $t=1$, shown in Fig.\,\ref{fig:Model}(c), $\sigma$ transforms as
	\begin{equation}
		\sigma \mapsto  R \sigma R^T
	\end{equation}
	\normalsize
	where $R=S_k^{(d)}\cdots S_{0}^{(i+1)}S_{1}^{(i)}\cdots S_{0}^{(2)}S_{1}^{(1)}$, 
	the index $k=\mathrm{mod}_{2}(d)$ depends on the depth $d$ of the array, and the superscript $i$ denotes the $i$th layer. The matrix representations of the even- and odd-numbered layers are $n \times n$ block matrices denoted by $S_0^{(i)}$ and $S_1^{(j)}$ respectively. These block matrices correspond to the symplectic transformations associated with the individual OPAs. These symplectic transformations are derived from the Bogoliubov transformations associated with the Hamiltonian. See Appendix A for details.
	
    \emph{Computational complexity.---} It has been well established that sampling the Boson distribution from the output of a linear interferometer network with Fock and Gaussian state as inputs is related to the computation of Permanent and Hafnian function, respectively\,\cite{aaronson2010computational,Hamilton_2017}. 
	As both functions are in the computational complexity class $\#P$ under certain computational theoretic assumptions. This underpins the use of BS as a candidate for demonstrating quantum computational advantage

	The hardness of our system can be shown using the Hafnian master theorem\,\cite{KOCHAROVSKY2022144}, which is valid for any $2n \times 2n$ symmetric matrix. Therefore, the covariance representation $R\sigma R^T$ of the output state is directly translated into a joint probability distribution for measuring a given sample of photons $\{n_k\}_{k=1}^{m}$,
	\begin{equation}
		p\left(\{n_k\}\right)=\frac{\mathrm{haf}\,\,\tilde{W}\left(\{n_k\}\right)}{\sqrt{\mathrm{det}(I+G)}\underset{k}{\prod}n_k!}
	\end{equation}
	where $G=R\sigma R^T$, and $\tilde{W}\left(\{n_k\}\right)$ is a $\sum_k{n_k} \times \sum_k{n_k}$ block matrix built from $W=\begin{bmatrix}
		0 & I\\ 
		I & 0
	\end{bmatrix}G(G+I)^{-1}$. The matrix $\tilde{W}$ is constructed as follows: for $1 < i,j \leq m$, entry $W_{ij}$ is replaced with an $n_i \times n_j$ block matrix filled with entries $W_{ij}$. The matrices $W_{i,j+m}$, and $W_{i+m,j}$ are constructed similarly. 
	
	%
   By the Bloch--Messiah decomposition\,\cite{BlochMessiah1962, CariolaroPierobon2016}, a lossless Gaussian unitary of this kind is equivalent to the standard GBS architecture consisting of passive linear optics acting on single-mode squeezed inputs. Therefore, in the lossless case our proposal is not a new complexity model distinct from standard GBS; rather, it is an alternative physical realization of the same class of Gaussian states. The general case including loss remains an open question.
    
    The same complexity restrictions that apply to linear GBS also extend to our scheme. Currently, the fastest algorithm not based on tensor network is $O(n^3 2^{\frac{n}{2}})$ \,\cite{björklund2019fasterhafnianformulacomplex} where $n$ is the dimension of the matrix. Conversely, the scaling of entanglement with $r$ and $d$ indicates the viability of classical simulation using tensor network, as this scaling governs how quickly the output statistics become computationally intractable to compute.
	
	\emph{Entanglement.---}The entanglement of a multimode Gaussian state can be estimated via the logarithmic negativity\,\cite{PhysRevLett.95.090503,PhysRevLett.90.047904}. The output is partitioned into two subsystems $A$ and $B$. The covariance matrix undergoes the following transformation under the partial transposition operation:
	\begin{equation}
		\sigma \mapsto \widetilde{V}=(I \oplus I_\text{mod} ) \sigma (I \oplus I_\text{mod})
	\end{equation}
	where $I_\text{mod}=\oplus_{i=1}^{n} Z_{\mathrm{Pauli}}$ with $Z_{\mathrm{Pauli}}$ the $z$ pauli matrix. The dimension of $I_\text{mod}$ depends on the partitions $A$ and $B$. Then, the logarithmic negativity can be expressed using the symplectic eigenvalues $\{\widetilde{\nu}_k\}$ of the matrix $\widetilde{V}$ with $\widetilde{\nu}_k>0$,
	\begin{equation}
		\label{entanglementideal}
		E_\mathcal{N}(\widetilde{V})=\sum_{k}F(\widetilde{\nu}_k)
	\end{equation}
	where $F(x)=\mathrm{max}\{0,-\log(x)\}$. 
	
	\begin{figure}
		\includegraphics[scale=0.6]{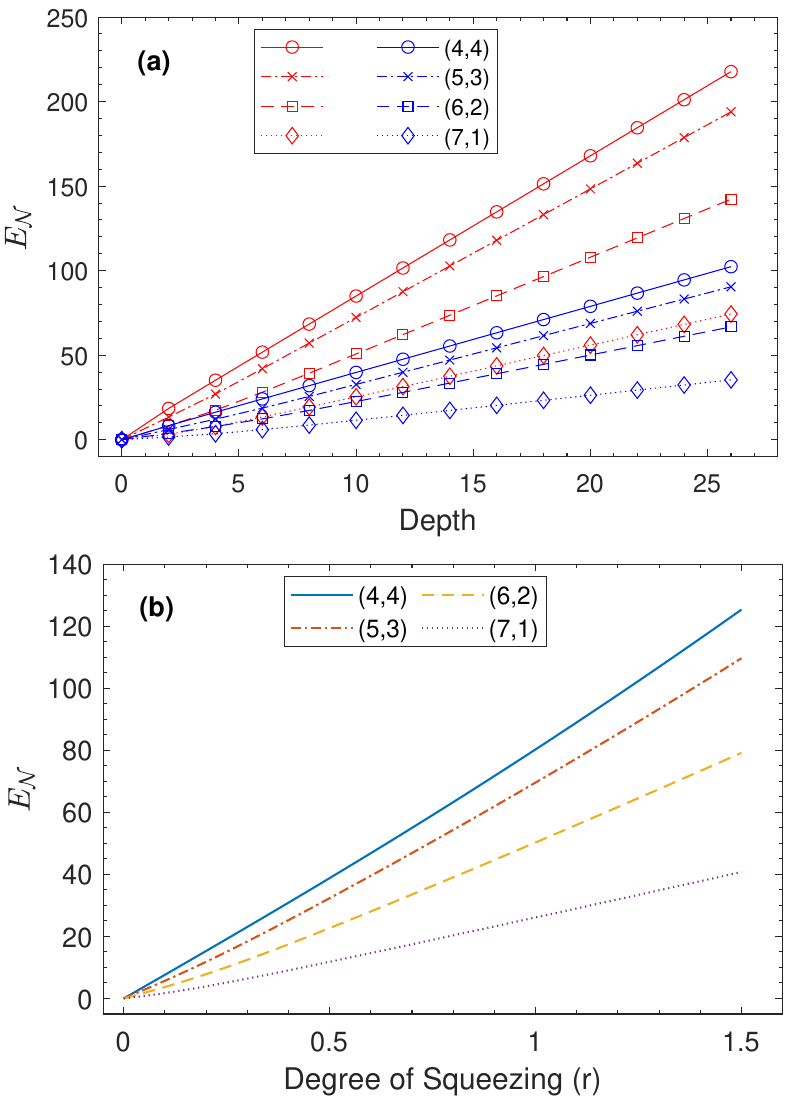}
		\caption{ Logarithmic negativity ($E_\mathcal{N}$) of the output state from an $8$ modes nonlinear interferometer network with vacuum input. The degree of squeezing is uniform across all OPAs. (a) shows $E_\mathcal{N}$ as a function of depth $d$ for two different degrees of squeezing $r=0.8$ (blue) and $r=1.6$ (red). The circle-solid, cross-dashed, square-dotted and diamond-dotted-dashed lines correspond to the four partition strategies: (4,4), (5,3), (6,2) and (7,1), respectively. (b) shows the corresponding $E_\mathcal{N}$ over the degree of squeezing with a typical range from 0 to 3. The evolution depth is set to $16$ to fully connect all modes.
		}
		\label{EntanglmentNoLoss}
	\end{figure}
	To demonstrate the entanglement growth, we numerically analyzed $E_\mathcal{N}$ in the case where nothing appearing in the input channels, that is, the scenario with vacuum states as inputs. As anticipated, all the entanglement originates from the nonclassicality of the multimode light field. The degree of entanglement depends on both the number of squeezers and the squeezing strength. Intuitively, increasing either parameter enhances the entanglement injected into the system.
 
	We consider an active network composed of a fixed number of $8$ input-output modes. To investigate the entanglement, the output multimode state can be grouped using four partitions: $(4,4)$, $(5,3)$, $(6,2)$, $(7,1)$. For instance, if the modes are labeled as $(1,2,3,4,5,6,7,8)$, the partition $(5,3)$ divides the system into two subsystems $(1,2,3,4,5)$ and $(6,7,8)$. Other partitions are defined analogously. 
    As shown in Fig.\,\ref{EntanglmentNoLoss}(a), for two typical squeezing parameters $r=0.8$ and $r=1.6$, $E_\mathcal{N}$ exhibits a linear relationship with the evolution depth $d$ once $d$ is sufficiently large i.e., when full mode connectivity is achieved. This linear growth is observed across all partition strategies.
	Similar to conventional GBS with passive elements, an equal partition yields the highest entanglement as illustrated by the solid line with circles in Fig.\,\ref{EntanglmentNoLoss}(a). 
	For a given partition strategy, the entanglement grows with increasing squeezing.
	Fig.\,\ref{EntanglmentNoLoss}(b) shows $E_\mathcal{N}$ as a function of the squeezing parameter $r$ at fixed depth of $16$. The entanglement measure $E_\mathcal{N}$ demonstrates an asymptotic linear dependence on the squeezing parameter $r$, with the rate of increase being partition-dependent. Notably, the equal partition $(4,4)$ produces the steepest increase.
	Thus, our scheme, without considering loss, can generate an entangled multimode Gaussian state with a tunable degree of entanglement that is linear with respect to $r$ and $d$. 

 \begin{figure*}[!ht]
	\includegraphics[width=500px]{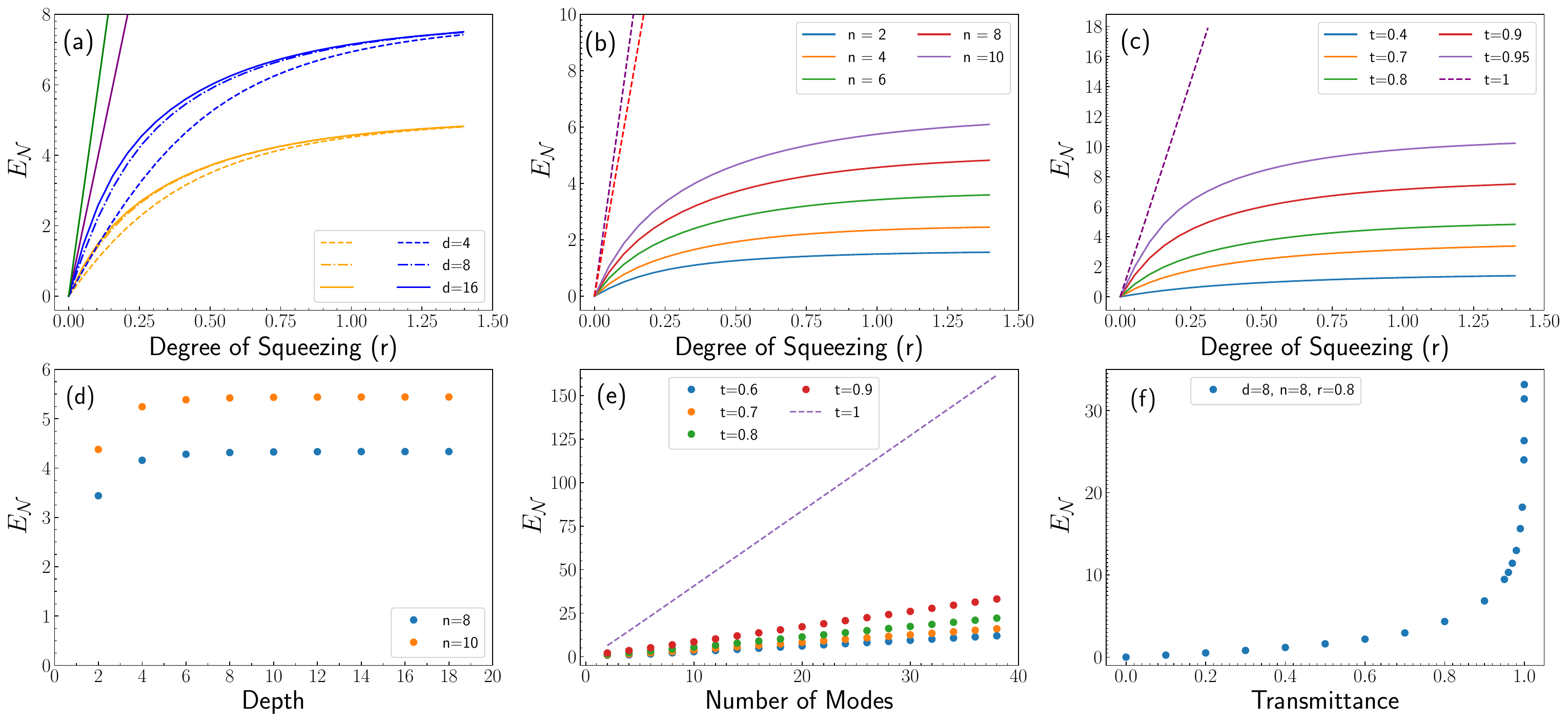}
	\caption{Logarithmic negativity ($E_\mathcal{N}$) of the output states in the presence of photon loss. a) $E_\mathcal{N}$ varying with the degree of
		squeezing ($r$) is shown for 8-mode vacuum inputs with transmission rate ($t$) of $0.8$ (yellow) and $0.9$
		(blue) at a depth ($d$) of $4$, $8$, and $16$ corresponding to the dash, point dash, and solid line. For reference, the purple and green solid lines show the ideal case for $d=8$ and $d=16$ respectively.  b) A comparison of $E_\mathcal{N}$ varying with the degree of squeezing for $n =2$, $4$, $6$, $8$, and $10$ modes
		with $t=0.8$ and $d= 12$ . The red and purple dashed lines show the ideal case for $n=8$ and $n=10$ respectively. c) Similar to b), for various transmission with $8$-mode and $d = 12$. d)
		The scaling of logarithm negativity in terms of depth for $n=8$, and $10$ modes with $t = 0.8$ and $r = 0.8$ e) The scaling of $E_\mathcal{N}$ with the number of modes is shown with $d = 8$ and $r = 0.8$ for  $t=0.6$, $0.7$, $0.8$, and $0.9$ (solid-colored circles). The ideal scaling is shown as well for reference (dash line). f) The scaling of $E_\mathcal{N}$ in terms of the transmission rate for $r = 0.8$, $d=8$, and $n = 8$.}
	\label{Entanglment_loss}
\end{figure*}
	
	\emph{Effect of photon loss.---} We have established that GBS with an $\mathrm{SU}(1,1)$ interferometer array is computationally challenging under ideal conditions. Before proceeding, we note that, in the lossless limit, the Bloch-Messiah decomposition\,\cite{PhysRevA.93.062115} shows that our nonlinear network is equivalent to standard GBS. When photon loss is introduced, this equivalence breaks down: while loss channels commute through $\mathrm{SU}(2)$ interferometers, they do not in an $\mathrm{SU}(1,1)$ network. See Appendix C for details. 
	With loss taken into consideration, an algorithm recently developed based on the theory of tensor network demonstrates good efficiency in simulating the results of the state-of-the-art GBS experiments 
	\,\cite{Oh2024,PhysRevA.104.022407}. However, such an algorithm does not invalidate the demonstration of quantum advantage in BS, as it remains unable to approximate the results of GBS when the entanglement scaling with the photon number is at least linear. 
	By examining $E_\mathcal{N}$, we show that it exhibits linear growth with the presence of loss in terms of number of modes. Thus, the computational hardness of our scheme persists even under realistic photon loss. 

	The case where the number of OPA layers is odd can be derived easily from the even case. Therefore, we assume $d$ is even and let $m=\frac{d}{2}$. The general transformation associated with two consecutive OPA layers $k$ and $k+1$,  including the beam splitters, can be expressed in a concise form: 
	\begin{equation}
		\begin{bmatrix}
			V_i & U_i & Q_i\\
			0 & I &0 \\
			0 & 0 &I \\
		\end{bmatrix}
		\begin{bmatrix}
			\vec{A}\\
			\vec{F}_k\\
			\vec{F}_{k+1}
		\end{bmatrix}
	\end{equation}
	
	\noindent where $V_{i}$ is a $2n \times 2n$ matrix encoding the combined transformation of the OPAs with loss on the input operators. $U_{i}$ is a $2n \times 2n$ matrix that introduce environment operators due to the beam splitters of the $k$th layer, similarly $Q_i$ corresponds to the beam splitter of the $k+1$st layer. The vector $\vec{A}= [\vec{a}, \vec{a}^{\dagger}]^T$  and  $\vec{F}_k=[\vec{f}_k,\vec{f}^{\dagger}_k]^T$ with $\vec{a}=(a_1,...,a_n)$, $\vec{a}^{\dagger}=(a^{\dagger}_1,...,a^{\dagger}_n)$, and $\vec{f}_k=(f_{k,1},...,f_{k,n})$. The operator $f_{k,i}$ is the annihilation operator of the environment due to the $k$th layer's $i$th beam splitter. The vector $\vec{f}^{\dagger}_{k+1}$ is similarly defined. The commutation relation for the environment operators is $[f_{k,i},f^{\dagger}_{l,j}]=\delta_{kl}\delta_{ij}$.
	Then, the resulting operators are 
	\begin{align}
		\begin{split}
				\vec{A}_{out} & = \prod_{i=1}^{m}V_i \vec{A} +\sum_{i=1}^{m} \prod_{j=i+1}^{m}V_j U_i \vec{F}_{2i-1} + \\ & \sum_{i=1}^{m} \prod_{j=i+1}^{m}V_j Q_i \vec{F}_{2i}^{\dagger}			
		\end{split} \label{main_res1}
	\end{align}            
    with $\prod_{j=m+1}^{m}V_j=I$. See Appendix B for details. 	From the above equation, the correlation functions can be found easily. For simplicity, while preserving the key considerations, we assume uniform squeezing $r$ across all OPAs and identical transmittance $t$ for all beam splitters. In addition, we set all squeezing angles $\theta=0$. 

As illustrated in Fig.\,\ref{Entanglment_loss}, we numerically calculate $E_\mathcal{N}$ under conditions of photon loss. The upper panels present the entanglement growth in response to an increasing squeezing parameter, examining scenarios with varying evolution depths (a), differing numbers of modes (b), and assorted transmittance rates (c). As expected, the rate of entanglement growth with increasing squeezing is suppressed by photon loss, causing the linear growth observed in the lossless case to degrade into a slower, sublinear increase. This indicates that a greater degrees of squeezing do not yield proportional gains in entanglement due to the concurrent amplification of vacuum noise caused by photon loss, and confirms that decreasing the photon loss rate  is more effective in preserving the complexity of GBS than increasing the initial degree of squeezing\,\cite{PhysRevA.108.052604}. Nonetheless, it is important to acknowledge that, regardless of transmittance, entanglement continues to grow with increased squeezing, albeit at a progressively slower rate.

In the ideal scenario, for a fixed number of modes,the entanglement scales linearly with the evolution depth, as shown in Fig.\,\ref{EntanglmentNoLoss}(a). However, in the presence of photon loss, as shown in Fig.\,\ref{Entanglment_loss}(a), $E_\mathcal{N}$ converges to a common scaling irrespective of the depth as $r$ increases. That is to say, simply increasing the evolution depth does not lead to continuous entanglement growth; instead, the system approaches a steady-state regime, as shown in Fig.\,\ref{Entanglment_loss}(d). We can also observe this from the single mode case. 
In a conventional optical parametric oscillator, the degree of squeezing in the generated single-mode squeezed state is determined solely by the gain and the total cavity loss\,\cite{Patera2010}. Our consideration can be taken as a generalization to the multimode, where the degree of entanglement for a given number of modes, which also reflects the nonclassicality of the multimode state, is determined only by the degree of squeezing and the rate of photon loss when the evolution depth is sufficiently large. This can be understood as a competition between amplifying the signal and the accumulated noise: nonclassical or entangled state can only be sustained when the photon loss for each depth is lower than the gain.  
 \begin{figure*}[!ht]
	\includegraphics[width=500px]{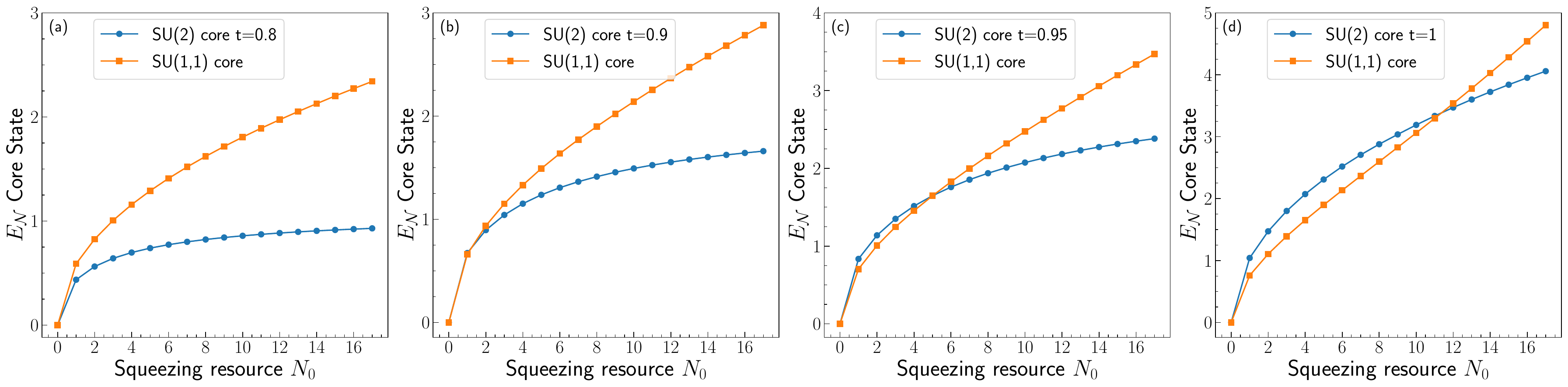}
	\caption{ Core state logarithmic negativity \(E_{\mathcal N}(V_p)\) in terms of squeezing resource \(N_0\) for \(8\) modes and \(4\) layers, shown for inter-layer transmission rates \(t=0.8\) (a), \(0.9\) (b), \(0.95,\) (c), and \(1\) (d). The plots compare the decomposed core entanglement of SU(2)-based (blue) and SU(1,1)-based (orange) Gaussian boson sampling architectures under the same squeezing budget \(N_0\).}
	\label{fig:ENmaxVp_grid}
\end{figure*}

To show the quantum advantage using GBS, the number of modes is one of the key parameters determining the system size besides the degree of squeezing. It has been shown that, with the presence of photon loss, the asymptotical entanglement entropy scaling with the number of modes is closely relates to classical simulation complexity\,\cite{PhysRevA.108.052604,PhysRevA.69.052308}. Ideally, this scaling is linear in the number of modes $n$. However, in a lossy linear interferometer, the entanglement growth can degrade significantly, and even decrease, making classical simulation of lossy GBS tractable.

In Fig.\,\ref{Entanglment_loss}(b), although photon loss suppresses entanglement growth with respect to the squeezing parameter, a sufficiently large degree of squeezing ensures that the final entanglement scales proportionally with the number of modes. This behavior is made explicit in Fig.\,\ref{Entanglment_loss}(e) which demonstrates the linear scaling of our boosted GBS under realistic levels of transmittance, as found in experiments. While any photon loss will significantly decrease the growing rate of resulting entanglement of the output state when compared to the lossless scenario (purple dashed line), the linear scaling between the entanglement and number of modes persists even for large degree of photon loss.

This is a key finding of our work: the reduction in simulation complexity observed in conventional linear GBS under photon loss is not present in our boosting scheme. By utilizing a nonlinear interferometer network composed of synchronously pumped optical parametric amplifiers (OPAs), quantum advantage can still be demonstrated in non-ideal, lossy environments. In contrast to the experimental difficulty of minimizing photon loss in linear interferometers, our scheme offers an alternative that is compatible with existing squeezed-light platform. Squeezed states are highly sensitive to photon loss as shown in Fig.\,\ref{Entanglment_loss}(f), even small losses significantly reduce entanglement. Without an inline injection of nonclassicality, only enhancing the degree of squeezing at the input alone cannot overcome the detrimental effects of photon loss.

\emph{Core decomposition and intrinsic entanglement comparison.---} 
The scaling of the full output-state entanglement is a useful diagnostic, but for noisy
Gaussian states it is also important to separate intrinsically quantum correlations from the
classical Gaussian noise that can be represented as random displacements. To make this
comparison more precise, given two output Gaussian states produced by two architectures under comparable loss conditions,
\[
V_1 \quad\text{(SU(2) output)},\qquad
V_2 \quad\text{(SU(1,1) output)},
\]
we want to compare the \emph{intrinsic} entanglement resource.  For each output covariance matrix $V$ (either $V_1$ or $V_2$), we seek a decomposition
\begin{equation}
	V = V_p + W,
	\label{eq:decomp}
\end{equation}
where $V_p$ is the covariance matrix of the \emph{quantum core} and must itself be physical, while $W \succeq 0$ is interpreted as a Gaussian random-displacement noise. Since the decomposition is generally not unique, it is usefull to  define a canonical core by minimizing the mean photon number of the core, which is equivalent to minimizing a trace objective in standard covariance-matrix conventions \cite{Oh2024}.

For a zero-mean Gaussian state, the total mean photon number is, up to a convention-dependent constant shift, a linear function of $\Tr(V)$. Therefore the optimization can be written as
\begin{equation}
	\min_{V_p,W}\ \Tr(V_p),
	\label{eq:objective}
\end{equation}
subject to the constraints
\begin{align}
	&V = V_p + W, \label{eq:constraint_decomp}\\
	&W \succeq 0, \label{eq:constraint_W}\\
	&V_p + i\Omega \succeq 0. \label{eq:constraint_Vp_phys}
\end{align}
Equations \eqref{eq:constraint_W}--\eqref{eq:constraint_Vp_phys} ensure that $W$ is a valid classical covariance and that $V_p$ is a valid quantum covariance matrix. This choice is intentionally conservative: it minimizes the quantum resource retained in the core.

Hence, for each $V \in \{V_1,V_2\}$, one solves a semidefinite program and denotes the optimizer by $(V_p,W)$. This yields the core covariance matrices
\[
V_{p,1} \ \text{from}\ V_1,\qquad
V_{p,2} \ \text{from}\ V_2.
\]
This procedure is architecture-agnostic: it depends only on the final covariances $V_1$ and $V_2$, so it applies equally to passive SU(2) networks and to SU(1,1) networks with interleaved vacuum loss.
 \begin{figure*}[!ht]
	\includegraphics[width=500px]{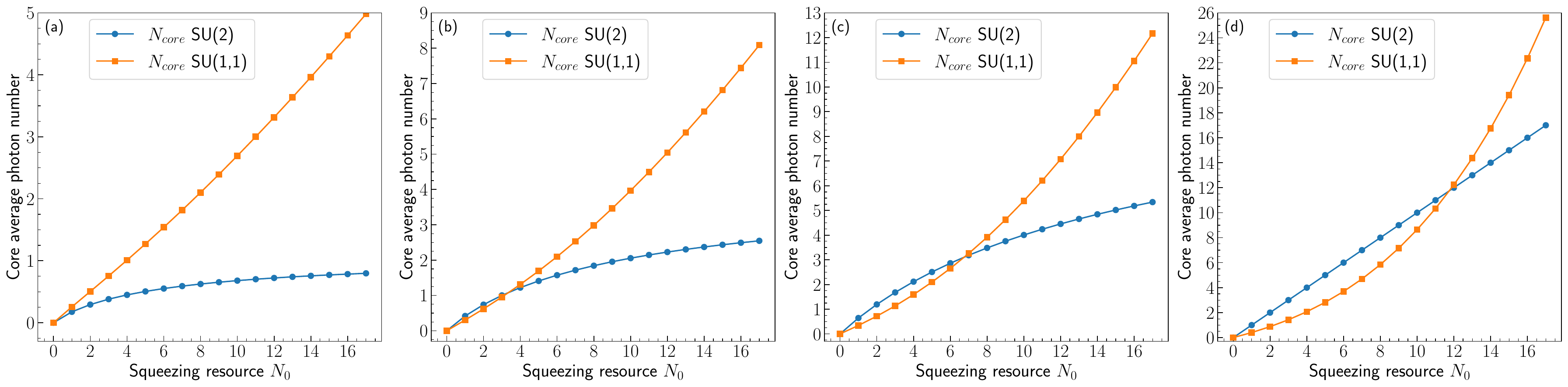}
	\caption{Core-state averate photon number as a function of squeezing resource \(N_0\) for \(8\) modes and \(4\) layers, shown for inter-layer transmission rates \(t=0.8\) (a), \(0.9\) (b), \(0.95,\) (c), and \(1\) (d). The plots compare the decomposed core entanglement of SU(2)-based (blue) and SU(1,1)-based (orange) Gaussian boson sampling architectures under the same squeezing budget, \(N_0\).}
	\label{fig:core_photons}
\end{figure*}

\emph{Comparing the extracted cores.---}
Once the cores are obtained, we evaluate logarithmic negativity,
\begin{equation}
	E_{\mathcal N}\!\left(V_{p,i};A|B\right),\qquad i\in\{1,2\}.
\end{equation}
We then compare the maximum
\begin{equation}
		E_{\mathcal N}\!\left(V_{p,1}\right)
		\ \text{and}\
		E_{\mathcal N}\!\left(V_{p,2}\right)
\end{equation}\label{eq:core_ent_compare}as a measure of the \emph{intrinsic} entanglement resources of the two architectures under the same extraction rule.

To ensure a fair comparison between SU(2)-based and SU(1,1)-based GBS architectures, we define the squeezing budget as the total mean number of photons generated by all active squeezing elements acting on vacuum in the \emph{lossless} device. Denoting this budget by $N_0$, we impose
\[
N_{\mathrm{sq}}^{(\mathrm{SU2})} = N_{\mathrm{sq}}^{(\mathrm{SU11})} = N_0.
\]
This definition is hardware-agnostic and tracks the nonlinearity or pump resource independently of how much energy is later removed by interleaved loss.

Our numerical results show that, under the same squeezing budget and the same loss model, the behavior of the core entanglement is different in the SU(2) and SU(1,1) architectures. For the passive SU(2) network, the logarithmic negativity of \(V_p\) exhibits clear saturation as the squeezing resource is increased, and this saturation becomes more pronounced as the inter-layer transmission decreases, as shown in Fig. \ref{fig:ENmaxVp_grid}. 

By contrast, for the active SU(1,1) network, the logarithmic negativity of the core state continues to grow substantially with the squeezing resource even in the presence of loss. Physically, this reflects the fact that in a passive SU(2) architecture the nonclassical correlations are created at the input and are subsequently degraded by loss, whereas in the SU(1,1) architecture the distributed OPAs can generate and partially restore correlations throughout the circuit after intermediate loss events. 

Therefore, the observed robustness of the SU(1,1) scheme is not simply due to a large noise contribution \(W\) inflating the entanglement of the full mixed state \(V\); rather, a significant and growing amount of entanglement remains in the effective pure core \(V_p\) itself. The decomposition analysis shows that the advantage of the SU(1,1) architecture persists at the level of the core state.

Fig. \ref{fig:core_photons} shows that the core-state photon number increases with squeezing in both architectures, but substantially more rapidly in the SU(1,1) case as the system approaches the lossless limit. The behavior is similar to the entanglement of the core state. The cross over is earlier at higher loss. The nonlinear network is able to sustain growth while the linear network plateaus rapidly. Although this does indicate hardness, it does show the SU$(1,1)$ provides more nonclassical ressources that its SU$(2)$ counterpart. This is the advantage of gradually injecting nonclassicallity. 

In standard fixed-$k$ boson sampling, one often focuses on a collision-free regime. The bosonic ``birthday paradox'' suggests that collisions become likely when the typical detected photon number $k$ approaches $\mathcal O(\sqrt n)$. 

However, for a general Gaussian Boson sampling device, collision events are naturally part of the output distribution, so collision-free operation is not a prerequisite for defining the sampling task. Rather, the collision-free or dilute regime is often invoked when discussing complexity-theoretic hardness. In addition, hardness results have also been established in the constant-collision regime \cite{GrierQuantum2022}, whereas the complexity of the fully high-collision regime is not yet fully established.

From an experimental perspective, collision events do not invalidate the data. If threshold detectors are used, collisions lead to ambiguity because multiple photons in one mode produce the same click outcome as a single photon. By contrast, with photon-number-resolving detectors, these outcomes are distinguished explicitly. In that sense, the birthday-paradox condition is not needed experimentally to make the data meaningful. GBS with photon-number-resolving detection is the original Fock-basis formulation of the model, and its probabilities are naturally defined for general occupation patterns, including collision events. 

In GBS, collisions are generally more prominent because the total photon number fluctuates and may be large. This means that the usual dilute-regime hardness arguments and simple collision-based intuition are no longer sufficient, so experiments rely on more sophisticated validation methods such as Bayesian \cite{Bentivegna2014BayesianValidation,Deng2023Jiuzhang3}, correlation-based\cite{Phillips2019GBSCorrelators}, HOG \cite{Boixo2018CharacterizingQS, Zhong2020QCAUsingPhotons}, or cross-entropy-type tests \cite{MartinezCifuentes2024LXEGBS}.

\begin{figure}[h!]
	\centering
	\includegraphics[width=0.9\linewidth]{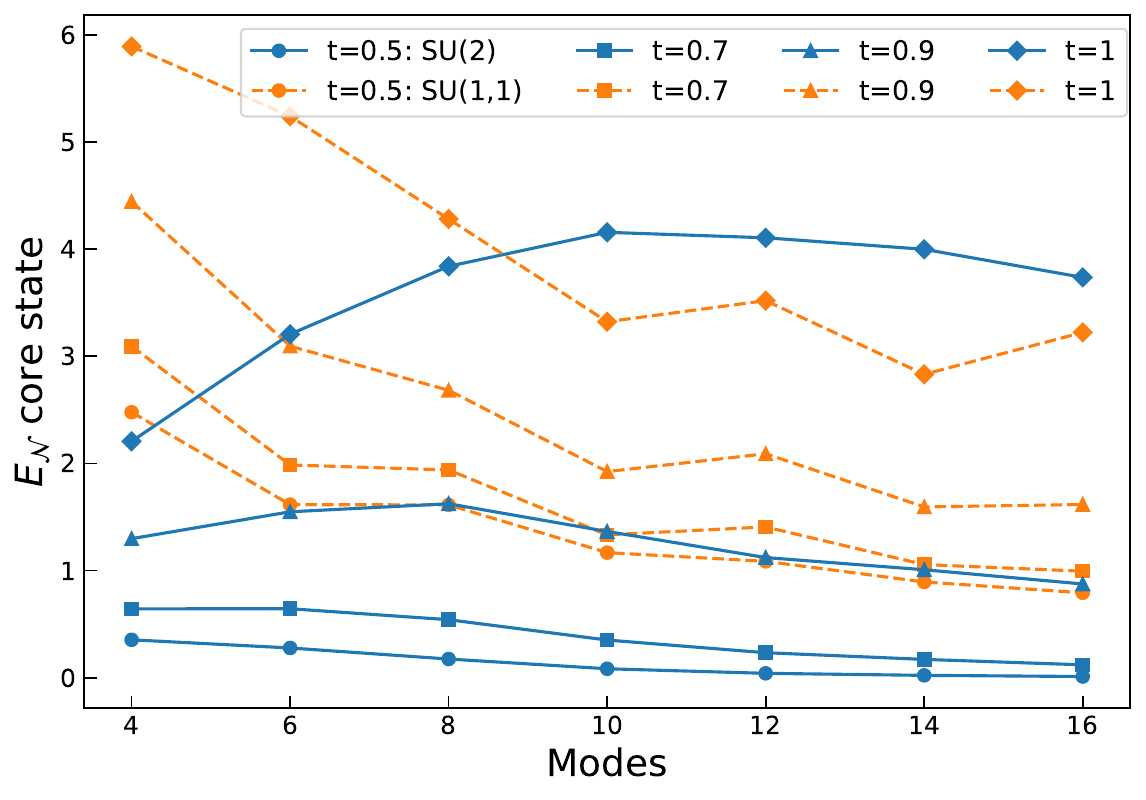}
	\caption{Core state logarithmic negativity \(E_{\mathcal N}(V_p)\) with different number of modes. The number of layeres is half of the number of modes. The graphs show for the logarithmic negativity for inter-layer transmission rates \(t=0.5\) (cicle), \(0.7\) (square), \(0.9\) (triangle), and \(1\) (diamond). The plots compare the decomposed core entanglement of SU(2)-based (blue) and SU(1,1)-based (orange) Gaussian boson sampling architectures under the same squeezing budget \(N_0=16\).}
	\label{fig:ENmax_N=16}
\end{figure}

With depth scaling as \(d \sim n/2\), to ensure full connectiviy, as shown in Fig \ref{fig:ENmax_N=16}, given a fixed squeezing budget, \(N_0=16\), and the same mode numbers, depth, and transmittivity, SU$(1,1)$ consistently outperform SU$(2)$ architecture in the presence of loss. Although it may perform worse in the lossless case for fixed total squeezing budget, the cause is rooted in \emph{where} and \emph{when} the nonclassical resources are generated within the network.

The scaling in terms of number of modes in the lossless case, the comparison is governed mainly by resource allocation. For an SU(2)-based Gaussian boson sampling architecture, squeezing is prepared at the input and the subsequent interferometer is passive. Thus, once the squeezed resource is injected, increasing the number of layers to achieve full connectivity does not consume additional squeezing. The passive network merely redistributes the existing nonclassical resource among the modes. By contrast, in an SU(1,1) architecture, the entangling elements are the optical parametric amplifiers themselves. If the total squeezing budget \(N_0\) is held fixed while the number of modes \(n\) increases and the depth scales as \(d \sim n/2\), to ensure full connectiviy, then the same finite resource must be distributed over an increasing number of active elements. As a result, the effective gain per OPA decreases with system size, and the correlations generated by each active layer become weaker. In that regime, the SU(1,1) network therefore show a reduction in logarithmic negativity as the number of modes increases.

 Once loss is introduced, however, the physical situation changes. In an SU(2) network, the squeezing is created only once, at the input, and then propagates through lossy channels and passive beam splitters. Loss removes photons and injects vacuum noise, thereby degrading the entanglement. Because the remainder of the network is passive, it cannot undo the effect of loss. The SU$(1,1)$ architecture behaves differently because it is active throughout the circuit. Its OPAs are distributed across the network, so quantum correlations are not generated only at the beginning, but repeatedly at multiple stages. If a lossy layer reduces the existing entanglement, a subsequent OPA can amplify the surviving signal and generate new pair correlations. In other words, the SU(1,1) network can continuously replenish its nonclassical resources after partial degradation. This makes SU$(1,1)$ intrinsically more robust against loss than an SU(2) network.

\emph{Discussion and Conclusion.---} We have proposed a scheme to enhance Gaussian Boson Sampling (GBS) using a nonlinear interferometric network composed of synchronously pumped optical parametric amplifiers (OPAs). By analyzing the logarithmic negativity $E_{\mathcal N}$ of the output covariance matrix, we show that the interferometer generates an entangled multimode Gaussian state whose entanglement scales linearly with both the squeezing parameter and the network depth. According to the Hafnian master theorem, sampling the photon-number distribution from this state remains $\#P$-hard, preserving the computational intractability of GBS.

Current GBS implementations are fundamentally limited by photon loss, which reduces both entanglement and computational complexity, allowing efficient classical simulation through tensor-network algorithms. 
Numerical simulations of our nonlinear SU(1,1) architecture reveal that, although loss suppresses entanglement, high squeezing compensates for this degradation. We find that $E_{\mathcal N}$ continues to scale linearly with the number of modes even under lossy conditions, with loss only affecting the slope of this relation.
The comparison  $E_{\mathcal N}$ and photon number of $V_p$, obtained from the decomposition of the final output into the core state, shows that SU(1,1) network consistantly outperform SU(2) network.

This robustness indicates that our OPA-boosted GBS network sustains computational hardness where passive SU(2) interferometers fail. Moreover, since the input need not be restricted to Gaussian states, the noncompact SU(1,1) structure may enable sampling problems of even greater complexity. These findings identify synchronously pumped OPA networks as a promising route toward scalable, loss-tolerant photonic quantum advantage.

We also emphasize that the present work addresses loss tolerance at the theoretical level. A scalable OPA network would require stable relative phases between the pump fields and the generated signal and idler modes, as well as temporal and spectral synchronization of the pump pulses across the active elements. The associated phase-locking loops, pump distribution, mode matching, and timing control would introduce additional hardware overhead and possible error channels. These requirements are familiar from synchronously pumped optical-parametric-oscillator and large-scale continuous-variable experiments~\cite{Patera2010,Averchenko2011,Yokoyama2013,Yoshikawa2016,Cialdi2021}. Therefore, the scalability suggested by our loss model should be understood as conditional on maintaining phase coherence and synchronization throughout the network.

\begin{acknowledgments}
	The authors acknowledge insightful comments from Liang Jiang, Zheng-Hao Liu, Changhun Ou, Martin B. Plenio and Man-Hong Yung.
	This work was supported by Innovation Program for Quantum Science and Technology (Nos.\,2021ZD0301200), National Natural Science Foundation of China (Nos.\,12474494, 12204468), China Postdoctoral Science Foundation (No.\,2024M753083), National Postdoctoral Program for Innovative Talents (Grant No.\,BX20240353), Fundamental Research Funds for the Central Universities (No.\,WK2030000081) and Research and Development Program of Anhui
	Province (No.\,2022b1302007).
    The simulations presented in this work were performed using the \texttt{Strawberry Fields} photonic quantum computing library~\cite{Bromley_2020,Killoran_2019}.
\end{acknowledgments}

\section*{Data Availability}

The numerical data and code that support the findings of this work are openly available on GitHub~\cite{Zhao_OPA_GBS_GitHub}.

\bibliography{ref.bib}

	\newpage
	\onecolumngrid
  
	\newpage

	\appendix
	\section{Derivation of the Covariance Matrix Representation of the Output State Without Loss}\label{NoLoss}
	The Hamiltonian associated with an optical parametric amplifier (OPA) is $H=\hbar(\xi \hat{a}^{\dagger}\hat{b}^{\dagger}-\xi^{*}\hat{a}\hat{b})$, where $\xi=-re^{i\theta}$. The corresponding Bogoliubov transformation is    
	\begin{equation}	
		\begin{bmatrix}
			\hat{a}_{out}	\\
			\hat{b}_{out}^{\dagger}
		\end{bmatrix}
		=
		\begin{bmatrix}
			\cosh(r)& -e^{i\theta}\sinh(r)\\
			-e^{-i\theta}\sinh(r)& \cosh(r)\\
		\end{bmatrix}
		\begin{bmatrix}
			\hat{a}_{in}	\\
			\hat{b}_{in}^{\dagger}
		\end{bmatrix}
		\label{SUmatrix}
	\end{equation}\\
	Using the ordering $(a_1,a_2,...,a_n,a^{\dagger}_1,...,a^{\dagger}_n)$ and $(x_1,x_2,...,x_n,p_1,...,p_n)$ in the main text, the full Bogoliubov transformation \cite{agarwal_2012} for two-mode squeezing is  
	\begin{align}
		& g
		= 
		{\begin{bmatrix}
				\cosh(r) &  
				0  &   
				0 &   
				-e^{i\theta}\sinh(r)\\
				0 &   
				\cosh(r) &   
				-e^{i\theta}\sinh(r) & 
				0\\ 
				0 &   
				-e^{-i\theta}\sinh(r)&  \cosh(r) & 
				0 \\
				-e^{-i\theta}   \sinh(r) &   
				0 &   
				0 &   
				\cosh(r) \label{complexRep}
		\end{bmatrix} }
	\end{align}
	~\\
	Since we are working with Gaussian states, it is convenient to leverage the fact that such states are fully characterized by their first and second statistical moments. Moreover, the Hamiltonian governing OPA corresponds to a Gaussian unitary transformation. This makes the symplectic formalism particularly natural for our analysis. In this framework, we express all transformations using the symplectic group $\mathrm{Sp}(4,\mathbb R)$,  which acts linearly on the phase-space quadratures. The Bogoliubov transformation employed in \eqref{SUmatrix} is a $\mathrm{SU}(1,1)$ operation. We seek its $\mathrm{Sp}(4,\mathbb R)$ representation. This amounts to finding a transformation relating $(a_1,a_2,...,a_n,a^{\dagger}_1,...,a^{\dagger}_n)$ and $(x_1,x_2,...,x_n,p_1,...,p_n)$. A more detailed discussion can be found in \,\cite{2016gaussianquantummetrologyspacetime}. From the definition of quadrature operators $a_j=\frac{x_j+ip_j}{\sqrt{2}}$ and  $a^{\dagger}_j=\frac{x_j-ip_j}{\sqrt{2}}$, we find
	\begin{align}
		& T
		= \frac{1}{\sqrt{2}}
		{\begin{bmatrix}
				I &  iI\\
				I & -iI \label{xatransfrom}
		\end{bmatrix} }.
	\end{align}
	Using $T$, the Bogoliubov transformation $g$ can be transformed into an element in $\mathrm{Sp}(4,\mathbb R)$, resulting in 
	\begin{align}
		O =T^{\dagger}gT=
		{\begin{bmatrix}  
				\cosh(r)&  {-\cos(\theta)\sinh(r)}& 0&   -\sin(\theta)\sinh(r) \\
				-\cos(\theta)\sinh(r) &  \cosh(r) &  -\sin(\theta)\sinh(r)	&  0\\
				0&  -\sin(\theta)\sinh(r)  &   \cosh(r)&   \cos(\theta)\sinh(r)\\ 	-\sin(\theta)\sinh(r)	&  0&  \cos(\theta)\sinh(r)&  \cosh(r)\\ \label{SPrep}
		\end{bmatrix} },
	\end{align}
	satisfying $O\Omega O^T=\Omega$, with the entry of $\Omega$ defined as $[x_k,p_j]=i\Omega_{kj}$. The matrix $O$ can be generalized to a transformation on an $2m$-mode state, see \eqref{eq:ODD} and \eqref{eq:even} below. 
	
	\begin{align} &
		S_1=\begin{bmatrix}
			\begin{array}{c|c}
				\begin{matrix}
					A_{l1} & & \\
					& \ddots & \\
					& & A_{ln}
				\end{matrix}
				&
				\begin{matrix}
					B_{l1}& & \\
					&\ddots & \\
					& & B_{ln}
				\end{matrix}
				\\
				\hline
				\begin{matrix}
					B_{l1}& & \\
					&\ddots & \\
					& & B_{ln}
				\end{matrix}
				&
				\begin{matrix}
					C_{l1} & & \\
					& \ddots & \\
					& & C_{ln}
				\end{matrix}
			\end{array} 
		\end{bmatrix}\label{eq:ODD}
	\end{align}
	\begin{align}
		&
		S_0=\begin{bmatrix}
			\begin{array}{c|c}
				\begin{matrix}
					1&  & & \\
					& A_{l1} & & \\
					&   &\ddots  \\
					&   &  & A_{l,n-2} \\
					&   &  & & 1
				\end{matrix}
				&
				\begin{matrix}
					0&  & && \\
					&B_{l1}  & & \\
					&   &\ddots & \\
					&   & &B_{l,n-2}  &  \\
					&   &  & & 0 
				\end{matrix}
				\\
				\hline
				\begin{matrix}
					0&  & && \\
					&B_{l1}  & & \\
					&   &\ddots & \\
					&   &&B_{l,n-2}  &  \\
					&   &  & &0 
				\end{matrix}
				&
				\begin{matrix}
					1&  & & \\
					& C_{l1} & & \\
					&   &\ddots  \\
					&   &  & C_{l,n-2} \\
					&   &  & & 1
				\end{matrix}
			\end{array} 
		\end{bmatrix}\label{eq:even}
	\end{align}

	where the block matrices are defined as follows
	\begin{align*}
		&	A_{lj}
		=
		\begin{bmatrix}
			\cosh(r_{lj})& -\cos(\theta_{lj})\sinh(r_{lj}) \\
			-\cos(\theta_{lj})\sinh(r_{lj})& \cosh(r_{lj})\\
		\end{bmatrix} 
		\\
		&
		\\ &
		B_{lj}= \begin{bmatrix}
			0& -\sin(\theta_{lj})\sinh(r_{lj}) \\
			-\sin(\theta_{lj})\sinh(r_{lj})	& 0\\
		\end{bmatrix} 
		\\
		& 
		\\    & 
		C_{lj}= 
		\begin{bmatrix}
			\cosh(r_{lj})& \cos(\theta_{lj})\sinh(r_{lj}) \\
			\cos(\theta_{lj})\sinh(r_{lj})& \cosh(r_{lj})\\
		\end{bmatrix}
	\end{align*}
	where $l$ is the $l$th layer and $j$ for the $j$th OPA of the $l$th layer, $r_{lj}$ is the degree of squeezing and $\theta_{lj}$ is the phase.
	
	\noindent Once the symplectic representation of for each individual layer is found, the covariance matrix is transformed as
	\begin{equation}
		\sigma \mapsto  S_k^{(d)}\cdots S_{0}^{(i+1)}S_{1}^{(i)}\cdots S_{0}^{(2)}S_{1}^{(1)} \sigma \left(S_k^{(d)}\cdots S_{0}^{(i+1)}S_{1}^{(i)}\cdots S_{0}^{(2)}S_{1}^{(1)}\right)^T
	\end{equation}
	\normalsize
    with $k=d\mod{2}$.	
	
	\section{Covariance Matrix of the Output State of the Interferometer with Loss}\label{OutputLoss}
    In this section, we derive the general transformation of the annihilation and creation operators, which will be used to construct the covariance matrix representation of the output Gaussian state in our interferometer network. Assume, $d$, the number of layers, is even. Let us define the annihilation and creation operator vectors for the system as $\vec{a}=(a_1,a_2,...a_n)$, $\vec{a}^{\dagger}=(a_1^{\dagger},a_2^{\dagger},...a_n^{\dagger})$, and similarly, define the environment mode operators as $\vec{f}, \vec{f}^{\dagger}$. To incorporate the effect of photon loss into the overall transformation, we treat the environment modes as auxiliary vacuum modes coupled through beam splitters. Now, Eq.\, \eqref{complexRep} can be rewritten using block matrices defined as follows
	\[
	A_{new_{ij}}
	=
	\begin{bmatrix}
		\cosh(r_{ij})& 0 \\
		0 & \cosh(r_{ij})\\
	\end{bmatrix}, 
	B_{new_{ij}}= \begin{bmatrix}
		0& -e^{i\theta_{ij}}\sinh(r_{ij}) \\
		-e^{i\theta_{ij}}\sinh(r_{ij}) & 0\\
	\end{bmatrix},
	\]

	\noindent where $i$ denotes the $i$th layer and $j$ denotes the $j$th OPA. Using the matrices defined above, the general transformations on $\vec{a}^{\dagger}$ and $\vec{a}$ for each OPA layer can be constructed. For our proposed scheme, the general transformation will have one of the following forms:
	
	\[S_{new1}=\begin{bmatrix}
		\begin{array}{c|c}
			\begin{matrix}
				A_{new_{i1}} & & \\
				& \ddots & \\
				& & A_{new_{ij}}
			\end{matrix}
			&
			\begin{matrix}
				B_{new_{i1}}& & \\
				&\ddots & \\
				& & B_{new_{ij}} 
			\end{matrix}
			\\
			\hline
			\begin{matrix}
				B^{\dagger}_{new_{i1}}& & \\
				&\ddots & \\
				& & B^{\dagger}_{new_{ij}} 
			\end{matrix}
			&
			\begin{matrix}
				A_{new_{i1}} & & \\
				& \ddots & \\
				& & A_{new_{ij}}
			\end{matrix}
		\end{array} 
	\end{bmatrix},
	\\
	\\
	\\
	S_{new0}=\begin{bmatrix}
		\begin{array}{c|c}
			\begin{matrix}
				1&  & & \\
				& A_{new_{i1}} & & \\
				&   &\ddots  \\
				&   &  & A_{new_{ij}} \\
				&   &  & & 1
			\end{matrix}
			&
			\begin{matrix}
				0&  & && \\
				&B_{new_{i1}}  & & \\
				&   &\ddots & \\
				&   & &B_{new_{ij}}  &  \\
				&   &  & & 0 
			\end{matrix}
			\\
			\hline
			\begin{matrix}
				0&  & && \\
				&B^{\dagger}_{new_{i1}}  & & \\
				&   &\ddots & \\
				&   &&B^{\dagger}_{new_{ij}}  &  \\
				&   &  & &0 
			\end{matrix}
			&
			\begin{matrix}
				1&  & & \\
				& A_{new_{i1}} & & \\
				&   &\ddots  \\
				&   &  & A_{new_{ij}} \\
				&   &  & & 1
			\end{matrix}
		\end{array} 
	\end{bmatrix}
	\]
	where $S_{new1}$ and $S_{new0}$ are the representations for odd and even layers respectively. The transformation associated with a beam splitter layer transformations is
	
	\[ \label{beamsplitter}
	Beam_{k}
	=
	\begin{bmatrix}
		T_{k}&R_{k}\\
		-R_{k} & T_{k},
	\end{bmatrix}
	\]
	where $T_k=\mathrm{diag}\left(t_{k1},t_{k2},...,t_{kj}\right)$,   $R_k=\mathrm{diag}\left(\sqrt{1-t_{k1}^2},\sqrt{1-t_{k2}^2},\dots, \sqrt{1-t_{kj}^2}\right)$. The index $k$ and $j$ denote the $k$th layer and the $j$th beam splitter respectively. Each $t_{ki}$ denotes a transmittance value. Using block matrices, the transformations can be written compactly as follows
	\begin{equation} \label{Compact1}
		S_{new1}
		\begin{bmatrix}
			\vec{a}\\
			\vec{a}^{\dagger}
		\end{bmatrix}
		=
		\begin{bmatrix}
			M& N \\
			N^{\dagger} & M
		\end{bmatrix}
		\begin{bmatrix}
			\vec{a}\\
			\vec{a}^{\dagger}
		\end{bmatrix}
	\end{equation}
	\begin{equation} \label{Compact2}
		S_{new0}
		\begin{bmatrix}
			\vec{a}\\
			\vec{a}^{\dagger}
		\end{bmatrix}
		=
		\begin{bmatrix}
			M'& N' \\
			N'^{\dagger} & M'
		\end{bmatrix}
		\begin{bmatrix}
			\vec{a}\\
			\vec{a}^{\dagger}
		\end{bmatrix}
	\end{equation}
	\begin{equation}
		Beam_{k} 
		\begin{bmatrix}
			\vec{a}\\
			\vec{f}
		\end{bmatrix}
		=
		\begin{bmatrix}
			T_{k}&R_{k}\\
			-R_{k} & T_{k}
		\end{bmatrix}
		\begin{bmatrix}
			\vec{a}\\
			\vec{f}
		\end{bmatrix}
	\end{equation}
	Using the above transformations, we obtain the following:
	\begin{align}
		\begin{split}
		\vec{a} \mapsto  (T_{k+1}M'T_{k}M+T_{k+1}N'T_{k}N^{\dagger})\vec{a}+(T_{k+1}M'T_{k}N+T_{k+1}N'T_{k}M)\vec{a}^{\dagger}+(T_{k+1}M'R_{k})\vec{f}_k &\\ +(T_{k+1}N'R_k)\vec{f}^{\dagger}_k+R_{k+1}\vec{f}_{k+1}+R_{k+1}\vec{f}^{\dagger}_{k+1}
		\end{split}
	\end{align}
	
    The vector $\vec{a}^{\dagger}$ can also be found similarly. In matrix form, 
	
		\begin{equation}
		\begin{bmatrix}
			P &X &Y& Z & R_{k+1}& R_{k+1}\\	X^{\dagger}&P^{\dagger}&Z^{\dagger}&Y^{\dagger}& R_{k+1} & R_{k+1}\\	0&0&I&0&0&0\\
			0&0&0&I&0&0\\
			0&0&0&0&1&0\\
			0&0&0&0&0&1\\
		\end{bmatrix}
		\begin{bmatrix}
			\vec{a}\\
			\vec{a}^{\dagger}\\
			\vec{f}_k\\
			\vec{f}^{\dagger}_k\\
			\vec{f}_{k+1}\\
			\vec{f}^{\dagger}_{k+1}
		\end{bmatrix}\label{sample}
	\end{equation}
	where $P=T_{k+1}M'T_{k}M+T_{k+1}N'T_{k}N^{\dagger}$, $X=T_{k+1}M'T_{k}N+T_{k+1}N'T_{k}M$, $Y=T_{k+1}M'R_{k}$, and $Z=T_{k+1}N'R_k$ and $Q=R_{k+1}$. We can rewrite the above more concisely

		\begin{equation}
		\begin{bmatrix}
			V_i & U_i & Q_i\\
			0 & I &0 \\
			0 & 0 &I \\
		\end{bmatrix}
		\begin{bmatrix}
			\vec{A}\\
			\vec{F}_k\\
			\vec{F}_{k+1} \label{BASE}
		\end{bmatrix}
	\end{equation}
	where $\vec{A}=[\vec{a},\vec{a}^{\dagger}]^T$ and $\vec{F}_k=[\vec{f}_k,\vec{f}_k^{\dagger}]^T$.
		
	Since each layer introduce additional $\vec{f}$ operators, by modifying eq.\ref{BASE} appropriately by adding extra columns and rows, after $m$ layers, one can find 
	\begin{equation}
		\vec{A}_{out}= \prod_{i=1}^{m}V_i \vec{A} +\sum_{i=1}^{m} \prod_{j=i+1}^{m}V_j U_i \vec{F}_{2i-1}+\sum_{i=1}^{m} \prod_{j=i+1}^{m}V_j Q_i \vec{F}_{2i}^{\dagger}
	\end{equation}
	
	with $\prod_{j=m+1}^{m}V_j=I$. Eq.\ref{main_res1} in the main text is found. For vacuum environment, the expectation value is

	\begin{align}
		\begin{split}
		\label{expectationval}	\left\langle A_{out_l}A_{out_k}  \right\rangle & = \sum_{x=1}^{n}\left (\prod_{i=1}^{m}V_i\right )_{l,x}\left(\prod_{i=1}^{m}V_i\right )_{k,x+n} +\sum_{i=1}^{m} \sum_{x=1}^{n} \left(\prod_{j=i+1}^{m}V_j U_i\right)_{l,x} \left(\prod_{j=i+1}^{m}V_j U_i\right)_{k,x+n}\\& + \sum_{i=1}^{m} \sum_{x=1}^{n} \left(\prod_{j=i+1}^{m}V_j Q_i\right)_{l,x}\left(\prod_{j=i+1}^{m}V_j Q_i\right)_{k,x+n}
		\end{split}
	\end{align}
    With this the covariance matrix can be constructed.
	
\section{Inequivalence between Networks in a Lossy Environment}\label{notNetloss}

We show that when internal photon loss is present, a multimode network built from interleaved optical parametric amplifiers and beam splitters (loss) is \emph{not equivalent} to an network of the form ``one lumped loss+ lossless SU(1,1) network''.
The proof follows directly from Gaussian-channel composition rules.

A general $n$-mode Gaussian channel is specified by a pair of real matrices
\[
(X, Y),
\]
where $X, Y \in \mathbb{R}^{2n\times 2n}$ and $Y=Y^{\!\top}$, acting on first and second moments as
\begin{align}
\mathbf{d} &\longmapsto X\,\mathbf{d} + \mathbf{d}_0, \\
V &\longmapsto X\,V\,X^{\!\top} + Y.
\end{align}
For simplicity we set $\mathbf{d}_0 = 0$ in what follows.
\\
\\
The pair $(X,Y)$ must satisfy the complete positivity condition
\[
Y + i\Omega - iX\,\Omega\,X^{\!\top} \ge 0,
\]
where $\Omega = \bigoplus_{k=1}^{n}\!\begin{pmatrix}0 & 1 \\ -1 & 0\end{pmatrix}$ is the symplectic form.

\vspace{1em}
\noindent
\textbf{Composition rule.}
\\
Let two Gaussian channels $\Phi_1$ and $\Phi_2$ be represented by
\[
\Phi_1 = (X_1, Y_1), \qquad
\Phi_2 = (X_2, Y_2).
\]
Applying $\Phi_1$ first and then $\Phi_2$ gives the composed channel
\[
\Phi_{21} = \Phi_2 \circ \Phi_1,
\]
whose matrices $(X_{21}, Y_{21})$ are obtained as follows:
\begin{align}
X_{21} &= X_2 X_1, \label{eq:comp_X}\\[4pt]
Y_{21} &= Y_2 + X_2\,Y_1\,X_2^{\!\top}. \label{eq:comp_Y}
\end{align}

\noindent
\textit{Derivation.}
Starting from a covariance matrix $V$, the first channel yields
\[
V' = X_1 V X_1^{\!\top} + Y_1.
\]
Applying the second channel,
\[
V'' = X_2 V' X_2^{\!\top} + Y_2
     = X_2 X_1 V X_1^{\!\top} X_2^{\!\top}
       + (Y_2 + X_2 Y_1 X_2^{\!\top}),
\]
from which Eqs.~\eqref{eq:comp_X}--\eqref{eq:comp_Y} follow immediately.

\bigskip
\noindent
\textbf{Example.}  
A pure-loss channel with transmissivity $\eta$ has
\[
X_\eta = \sqrt{\eta}\,I_{2}, 
\qquad
Y_\eta = \tfrac{1-\eta}{2}\,I_{2}.
\]
Two sequential losses $\eta_1$ and $\eta_2$ compose as
\[
(X,Y) = (X_{\eta_2},Y_{\eta_2})\circ(X_{\eta_1},Y_{\eta_1})
      = (\sqrt{\eta_1\eta_2}\,I_{2},
         \tfrac{1-\eta_1\eta_2}{2}\,I_{2}),
\]
Changing the order results in the same result. This shows that losses commute and combine multiplicatively in transmissivity.
\\
\\
\begin{itemize}
\item \textbf{Passive SU(2) interferometer:} \((X,Y)=(R,0)\), with \(R\in\mathrm{Sp}(2n,\mathbb R)\cap \mathrm{O}(2n)\) (orthogonal symplectic).
\item \textbf{Single-mode squeezer:} on a single mode, the symplectic is
\[
S(r)=\begin{pmatrix} e^{r} & 0 \\[4pt] 0 & e^{-r}\end{pmatrix},
\]
and for that mode \((X,Y)=(S(r),0)\).
\item \textbf{Pure loss channel (vacuum environment):} for transmissivity \(0\le\eta\le1\),
\begin{equation}
X=\sqrt{\eta}\,I_{2},\qquad
Y=\frac{1-\eta}{2}\,I_{2},
\label{eq:loss}
\end{equation}
in the two-dimensional quadrature block for that mode. (For \(n\) modes with independent losses \(\eta_j\) this generalizes to the direct sum blocks.)
\end{itemize}
~\\
We now compare two possible orderings of loss and passive SU(2) transformation (beam splitter)
in a two-mode system. Because passive unitaries preserve total photon number and do not
squeeze quadratures, the noise transformations behave differently from the SU(1,1) case.
\\
~\\
Let the beam splitter be represented by a symplectic and orthogonal matrix
\[
R(\theta)=
\begin{pmatrix}
\cos\theta\ I & \sin\theta\ I\\
-\sin\theta\ I & \cos\theta\ I
\end{pmatrix},
\qquad
R(\theta)\,R(\theta)^{\mathsf T}=I.
\]
A uniform loss channel of transmissivity $\eta$ acts as
\[
(X_\eta,Y_\eta)=(\sqrt{\eta}\,I,\;\tfrac{1-\eta}{2}\,I).
\]
~\\
Case A: Loss after the interferometer.~\\
\begin{align}
& (X_1,Y_1)=(R(\theta),0), 
\qquad
(X_2,Y_2)=(\sqrt{\eta}\,I,\tfrac{1-\eta}{2}I),
\\[3pt]
& (X,Y)=(X_2,Y_2)\circ(X_1,Y_1)
=(\sqrt{\eta}\,R(\theta),\;\tfrac{1-\eta}{2}\,I).
\end{align}
~\\
Case B: Loss before the interferometer.~\\

\begin{align}
& (X_1,Y_1)=(\sqrt{\eta}\,I,\tfrac{1-\eta}{2}I),\qquad
(X_2,Y_2)=(R(\theta),0),
\\[3pt]
&(X,Y)=(X_2,Y_2)\circ(X_1,Y_1)
=(\sqrt{\eta}\,R(\theta),\;R(\theta)\,\tfrac{1-\eta}{2}I\,R(\theta)^{\mathsf T}).
\end{align}
Because $R(\theta)$ is orthogonal, $R(\theta)R(\theta)^{\mathsf T}=I$, we obtain
\begin{equation}
Y_{\text{after}}=\tfrac{1-\eta}{2}I, \qquad
Y_{\text{before}}=\tfrac{1-\eta}{2}I.
\end{equation}
Hence,
\[
(X,Y)_{\text{after}}=(X,Y)_{\text{before}},
\]
and the two orderings are \emph{equivalent} Gaussian channels.
\\
~\\
The key difference from the SU(1,1) (squeezing) case is that
a passive SU(2) transformation does not alter the isotropy of vacuum noise:
since it is an orthogonal rotation in phase space, it preserves
$Y\propto I$. Therefore, losses can be commuted through passive interferometers
and lumped into a single effective transmissivity,
\[
\eta_{\text{eff}} = \prod_j \eta_j.
\]
This property underlies the equivalence
\[
\text{Lossy SU(2) network} \;\equiv\;
\text{Global loss for each mode + Lossless SU(2) network .}
\]
~\\
The two-mode squeezing operation acts on the quadrature vector
$\bm{x}=(x_1,p_1,x_2,p_2)^{\mathsf T}$ as
\begin{equation}
S(r)=
\begin{pmatrix}
\cosh r\,I_2 & \sinh r\,Z\\
\sinh r\,Z & \cosh r\,I_2
\end{pmatrix},
\qquad
Z=\begin{pmatrix}1&0\\0&-1\end{pmatrix}.
\end{equation}

~\\
By the Bloch--Messiah (Euler) decomposition, any symplectic matrix can be written as
$S=O_1 D O_2$, where $O_{1,2}$ are orthogonal symplectic (passive) and $D$ is diagonal in single-mode squeezers.
For $S$, one finds
\begin{equation}
\boxed{
S(r)
= B_{50}\,
\big[S_1(r)\oplus S_2(-r)\big]\,
B_{50}^{\mathsf T},
}
\end{equation}
where
\begin{align}
B_{50} &= \tfrac{1}{\sqrt{2}}
\begin{pmatrix}
I_2 & I_2\\
-I_2 & I_2
\end{pmatrix},
&
S_1(r)\oplus S_2(-r)
&= \operatorname{diag}(e^{r},e^{-r},e^{-r},e^{r}).
\end{align}
Direct multiplication yields
\begin{equation}
S(r)=
\begin{pmatrix}
\cosh r\,I_2 & \sinh r\,Z\\
\sinh r\,Z & \cosh r\,I_2
\end{pmatrix},
\end{equation}
confirming the factorization.
\\
\\
Two-mode squeezer can be decomposed as passive $\rightarrow$ single-mode squeezing $\rightarrow$ passive. Hence, we focus on the single-mode squeezing. From the above, we know that beam splitter and loss commute. Hence, we focus on the single-mode squeezer and compare ``loss then squeeze'' vs.\ ``squeeze then loss'' for one mode. \\ \\

\noindent Case A: Squeeze then loss. \\ \\
Let \((X_1,Y_1)=(S(r),0)\) and \((X_2,Y_2)=(\sqrt{\eta}I,\tfrac{1-\eta}{2}I)\).
\begin{equation}
(X,Y)=(X_2X_1,\; Y_2 + X_2 Y_1 X_2^{\mathsf T})
=(\sqrt{\eta}\,S(r),\;\tfrac{1-\eta}{2}\,I).
\end{equation}

\noindent Case B: Loss then squeeze. \\ \\
Now \((X_1,Y_1)=(\sqrt{\eta}I,\tfrac{1-\eta}{2}I)\) and \((X_2,Y_2)=(S(r),0)\). Composition gives
\begin{equation}
(X,Y)=(S(r)\sqrt{\eta}I,\; 0 + S(r)\,\tfrac{1-\eta}{2}I\,S(r)^{\mathsf T})
=(\sqrt{\eta}\,S(r),\;\tfrac{1-\eta}{2}\,S(r)S(r)^{\mathsf T}).
\end{equation}
Since
\(
S(r)S(r)^{\mathsf T}=\operatorname{diag}(e^{2r},e^{-2r}),
\)
the noise matrices are
\begin{equation}
Y_{\text{after}}=\tfrac{1-\eta}{2}I, \qquad
Y_{\text{before}}=\tfrac{1-\eta}{2}\operatorname{diag}(e^{2r},e^{-2r}).
\end{equation}
They coincide for \(r=0\), and for $\eta=1$ the overall map is symplectic; Bloch–Messiah \cite{BlochMessiah1962, CariolaroPierobon2016} applies directly to the full network and it becomes the usual GBS. 
\\
\\
As two Gaussian channels are equal iff both \(X\) and \(Y\) coincide, the two orderings are inequivalent in general.

\[
\text{Lossy SU(1,1) network} \;\not\equiv\;
\text{Global loss for each mode + Lossless SU(1,1) network .}
\]

\section{Comparing Main Boson Sampling Scheme}

The scalability of standard Boson Sampling is limited because the success probability decreases exponentially with photon number, making high-$N$ events exceedingly rare. Optical loss further aggravates this problem and enables efficient classical simulation of noisy systems~\cite{PhysRevA.104.022407,Shchesnovich2021distinguishingnoisy,oh2023classicalsimulationalgorithmsnoisy}. Scattershot Boson Sampling provides a combinatorial enhancement in the probability of observing $N$-photon events, thereby improving scalability relative to the standard scheme. However, once loss is introduced, the same arguments that permit efficient classical simulation of noisy Boson Sampling remain applicable.
\\
~\\
Driven Boson Sampling further enhances the Scattershot protocol by achieving an $e$-fold increase in the effective input-state generation rate, thereby improving scalability. Although no experimental demonstration has yet been reported, it is expected that photon loss would similarly reduce this scheme to a classically simulable regime.
\\
~\\
Gaussian Boson Sampling (GBS) replaces single-photon Fock inputs with squeezed-vacuum states, making the system more robust to loss than previous schemes. As summarized in the table above, GBS has achieved major experimental progress. Nevertheless, tensor-network based methods such as matrix-product-state simulations can still efficiently approximate GBS output distributions under realistic loss.
\\
~\\
Our proposed approach introduces an active nonlinear interferometric network designed to enhance robustness against loss. The inclusion of nonlinearity ensures that tensor-network algorithms would require exponential computational resources even in lossy conditions, as evidenced by the linear scaling of entanglement with network shown in the main text. Consequently, our architecture potentially offers superior scalability compared to Gaussian Boson Sampling. Although, we mainly discuss squeezed vacuum state. It should be noted that a generic state, such as Fock state, could be used as well.

\begin{table*}[ht]
\centering
\renewcommand{\arraystretch}{1.25}
\setlength{\tabcolsep}{2pt}
\begin{tblr}{
  width = 0.97\textwidth,
  colspec = {|X[1.1]|X[1.0]|X[1.2]|X[1.1]|X[1.1]|X[1.0]|X[1.1]|X[1.3]|},
  row{1} = {font=\bfseries, bg=gray9!15},
  hlines, vlines
}
Scheme &
Source &
Network &
Detection &
Complexity &
Scalability &
Experiment &
Robustness to loss \\

Standard Boson Sampling~\cite{aaronson2010computational}& \SetCell[r=3]{} Heralded single photons  &
\SetCell[r=3]{} Passive linear network $\mathrm{SU}(2)$ &
\SetCell[r=2]{} PNR$^*$ &
\SetCell[r=3]{c} Permanents, \#P-hard &
Poor &
20-photon inputs ~\cite{PhysRevLett.123.250503} &
Classically simulable~\cite{PhysRevA.104.022407,Shchesnovich2021distinguishingnoisy,oh2023classicalsimulationalgorithmsnoisy} \\

Scattershot Boson Sampling~\cite{aaronson2013scattershotblog} &  & &
 &
 &
Slightly improved over standard BS &
12 SPDC sources ~\cite{PhysRevLett.121.250505} &
Classically simulable \\

Driven Boson Sampling~\cite{PhysRevLett.118.020502} &
&  & Threshold & &
Better than scattershot &
N/A &
Likely simulable under loss \\

\textcolor{blue}{Gaussian Boson Sampling}~\cite{Hamilton_2017} &
Squeezed vacuum states &
\textcolor{blue}{Passive linear} network $\mathrm{SU}(2)$ &
PNR\cite{Hamilton_2017} \newline threshold~\cite{PhysRevA.98.062322} &
Hafnians (PNR) \newline Torontonians (Thres.), \#P-hard &
\textcolor{blue}{Good} scalability &
Jiuzhang ~\cite{Zhong2020Jiuzhang1,Zhong2021Jiuzhang2,PhysRevLett.131.150601,liu2025robustquantumcomputationaladvantage}, Borealis ~\cite{Madsen2022} &
\textcolor{blue}{Can} be classically simulated using MPS/MPO~\cite{Oh2024,PhysRevA.108.052604} under realistic loss\\

\textcolor{red}{Proposed OPA-Boosted GBS} &
Squeezed vacuum states$^{**}$ &
\textcolor{red}{Nonlinear active} network $\mathrm{SU}(1,1)$ &
PNR &
Hafnian, \#P-hard &
\textcolor{red}{Potentially higher} scalability than GBS &
N/A &
\textcolor{red}{Cannot}  be classically simulated using MPS/MPO~\cite{Oh2024,PhysRevA.108.052604} under realistic loss \\
\end{tblr}
\caption{$^*$Photon Number Resolving detector. $^{**}$ Other state could be used. The table provide a comparison of major Boson Sampling schemes and  summarizes the main interest point within a Boson Sampling scheme.}
\end{table*}
\clearpage

\end{document}